\documentclass[prl,twocolumn,showpacs,showkeys,preprintnumbers,superscriptaddress,amsmath,amssymb]{revtex4-1}
\usepackage{graphicx}
\usepackage{longtable}
\usepackage{dcolumn}
\usepackage{bm}
\usepackage{xcolor}

\newcommand{\CuVS}{(Cu$_{2/3}$V$_{1/3}$)V$_2$S$_4$}

\begin{document}

\preprint{APS/123-QED}

\title{Fermi- to non-Fermi-liquid crossover and Kondo transition in two-dimensional \CuVS}

\author{A. Gauzzi}%
\email{andrea.gauzzi@upmc.fr}
\author{H. Moutaabbid}
\author{Y. Klein}%
\author{G. Loupias}%
\affiliation{Institut de Min\'eralogie, de Physique des Mat\'eriaux et de Cosmochimie (IMPMC) UMR7590, Sorbonne Universit\'e/CNRS/IRD/Museum National d'Histoire Naturelle, 4 place Jussieu, 75005 Paris, France}%
\author{V. Hardy}%
\affiliation{Laboratoire CRISMAT UMR 6508, CNRS/ENSICAEN/UCBN, Caen, France}%

\date{\today}

\begin{abstract}

By means of a specific heat ($C$) and electrical resistivity ($\varrho$) study, we give evidence of a pronounced Fermi liquid (FL) behavior with sizable mass renormalization, $m^{\ast}/m = 30$, up to unusually high temperatures $\sim$70 K in the layered system \CuVS. At low temperature, a marked upturn of both $C$ and $\varrho$ is suppressed by magnetic field, which suggests a picture of Kondo coupling between conduction electrons in the VS$_2$ layers and impurity spins of the V$^{3+}$ ions located between layers. This picture opens the possibility of controlling electronic correlations and the FL to non-FL crossover in simple layered materials. For instance, we envisage that the coupling between layers provided by the impurity spins may realize a two-channel Kondo state.    


\end{abstract}

\maketitle

                             
The solid state chemistry of layered transition metal chalcogenides (LTMC) displays ideal characteristics for tuning or even radically changing the electronic and transport properties of two-dimensional systems, which is promising for the development of novel materials in view of electronic, photonic and thermoelectric applications \cite{wan15}. Chemical intercalation has been widely used to electronically dope the conducting layers in a variety of LTMCs, such as 1T-TaS$_2$ \cite{wil75a} and 1T-TiSe$_2$ \cite{mor06}, which has unveiled an interesting interplay between charge density wave (CDW) orders and superconductivity (SC). Novel capabilities enabling the control of thickness down to the atomic limit and electronic doping \textit{via} ionic gating have led to striking discoveries, such as a transition from indirect to direct band gap in MoS$_2$ in the monolayer limit \cite{mak10} and high temperature SC in the FeSe system, where the critical temperature dramatically increases from 30 K in the bulk \cite{guo10} up to 100 K in monolayers \cite{ge14}.

In order to master the above properties, it is important to know to which extent band theory is applicable to LTMCs. In principle, one expects weaker electronic correlations in LTMCs than in transition metal (TM) oxides, considering the reduced ionicity of S and Se and the comparatively large bandwidth of hybrid S(3$p$) or Se(4$p$) - (TM)$d$ states. In practice, the question remains controversial even for well-studied compounds, such as the aforementioned 1T-TiSe$_2$ and 1T-TaS$_2$ systems. In Cu-intercalated 1T-TiSe$_2$, first-principles calculations within density functional theory explain the SC and CDW phases without invoking many-body effects \cite{cal11}, while photoemission experiments suggest that the Coulomb electron-hole interaction is required to stabilize the CDW phase \cite{cer07}. In 1T-TaS$_2$, it has been established that Mott localization controls the stability of a complex sequence of CDW and SC transitions \cite{wil75a,tho94}, as originally proposed by Fazekas and Tosatti \cite{faz79} and confirmed experimentally by various authors \cite{per06,sip08}. However, in spite of an intense research effort for more than forty years, the microscopic mechanism of these transitions remains elusive due to a complex interplay between electronic and lattice degrees of freedom \cite{yi18}.           

Here, we report on the observation of Fermi-liquid (FL) behavior with sizable mass renormalization and of a Kondo transition \cite{kon64,ste01} in the LTMC system \CuVS~ (CVS). To the best of our knowledge, this result has never been observed before in LTMCs. As shown in Fig. 5, CVS is characterized by a defect NiAs structure consisting of a stacking of 1T-VS$_2$ layers and of chains of edge-sharing (Cu$_{1-x}$V$_x$)S$_6$ octahedra \cite{kle11}. As compared to previous LTMCs, CVS displays no CDW or SC instabilities and no indications of strong electron-lattice coupling, which may explain why the FL to non-FL (NFL) crossover is clearly observed experimentally. 

The present finding is attributed to the peculiar properties of the above crystal structure enabling a coupling between conduction electrons in the layers and magnetic impurities in the chains. This coupling is the essential ingredient of the Kondo effect \cite{kon64} which gives rise to a singlet state with NFL properties. If confirmed, the picture proposed here would open the possibility of tuning the FL to NFL crossover in the regime of moderate electronic correlations characteristic of TM sulfides, a convenient regime for first-principles calculations. This would be a step towards a better understanding of correlated materials and their functionalization.     

Our original motivation was to verify the expectation of metallic transport in CVS, suggested by previous magnetic susceptibility data \cite{kle11} showing an enhanced Pauli-like term attributed to electronic correlations and no long-range magnetic order. For the present study, we have measured single crystals synthesized under high pressure, as described elsewhere \cite{kle11}. Typical crystal dimensions are 0.4 $\times$ 0.15 mm$^{2}$ in the $ab$-plane and 0.03 - 0.08 mm thickness along the $c$-axis. A few crystals were selected for dc resistivity measurements in a four-contact bar configuration using a commercial Quantum Design Physical Properties measurement system (PPMS). Specific heat measurements were carried out in the PPMS on a bunch of crystals from the same batch using a 2-$\tau$ relaxation method. Both types of measurements were carried out in the 2-300 K range by varying magnetic field up to 9 T. The reproducibility of the results was checked by measuring two batches of crystals grown under the same synthesis conditions.

\begin{figure}[ht]
\includegraphics[width=\columnwidth]{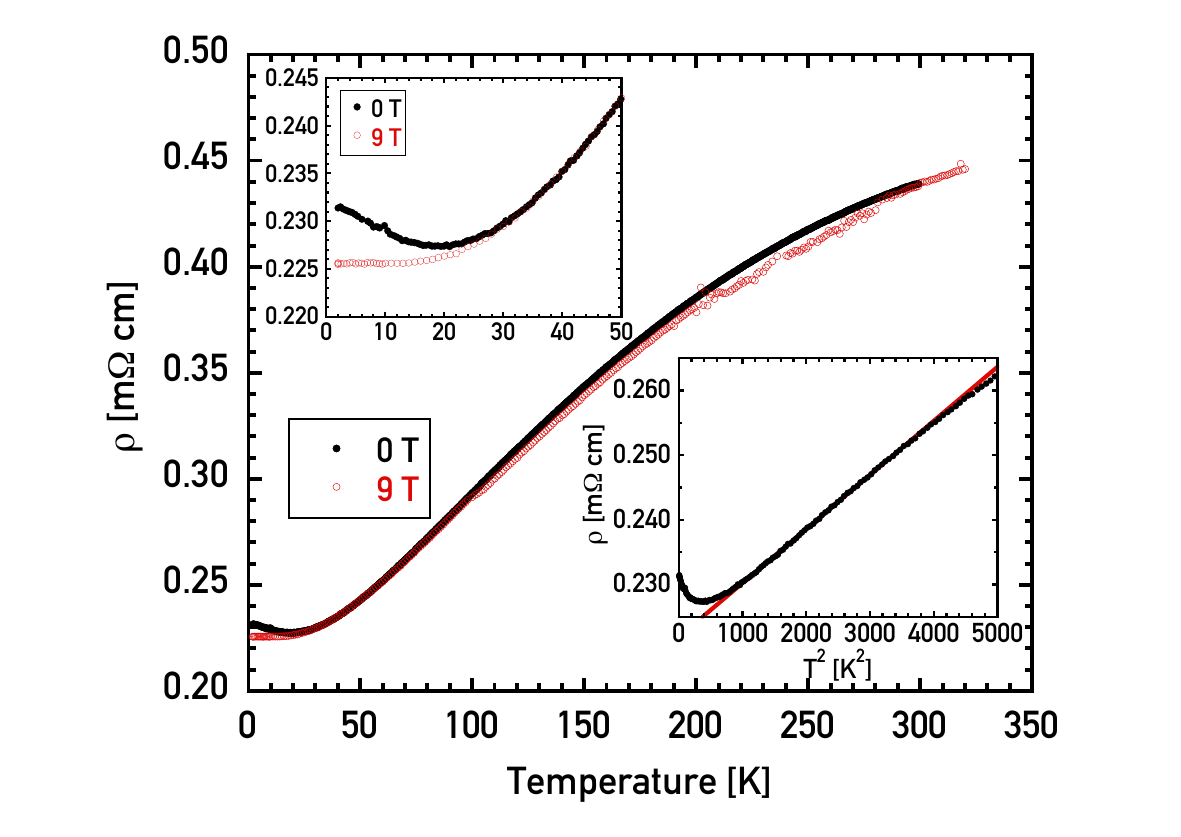}
\caption{(Color online) In-plane resistivity curves of a representative \CuVS~ single crystal taken at zero-field and at 9 T. Top inset: enlarged view of the minimum at $\sim$20 K suppressed by the field. Bottom inset: the linearity of the $\varrho$ vs. $T^2$ plot in the 30-70 K range gives evidence of Fermi-liquid behavior. The red solid line is a linear fit in this range. The results of the fit are given in the text.}
\label{fig1}
\end{figure}

Fig.~\ref{fig1} summarizes the results of the in-plane resistivity measurements. Note the following: (i) a marked metallic behavior with room temperature resistivity $\varrho_{\rm 300 K}$=0.44 m$\Omega$ cm and with a modest residual resistivity ratio, $RRR=\varrho_{\rm 300 K} / \varrho_{\rm 2K} \sim 2$; (ii) a pronounced upturn below $\sim$20 K completely suppressed by a magnetic field of 9 T. Feature (i) is characteristic of bad metals and common in 2D sulfides, such as V$_5$S$_8$ \cite{kaw75,noz75}, which shares with CVS a similar layered structure except it contains only 1/2 - instead of one - interstitial metal atom per VS$_2$ layer. (iii) In the temperature region between the upturn and 70 K, the $\varrho(T)$ curve displays a marked quadratic dependence, a signature of Fermi-liquid behavior arising from a dominant electron-electron scattering.

This is ascribed to the small Fermi energy characteristic of bad metals. To the best of our knowledge, a quadratic dependence of $\varrho(T)$ has been previously reported in transition metals \cite{ric68}, intermetallic compounds, such as A15 superconductors \cite{miy69}, TM oxides, such as V$_2$O$_3$ under high pressure \cite{mcw69} and Sr$_2$RuO$_4$ \cite{mae97}, and heavy-fermions \cite{kad86}, but never in LTMCs. A \textit{caveat} concerns a similar dependence reported in Ti$_{1+x}$S$_2$ \cite{tho75}, which shares with CVS a 1T-type layered structure with $x$ interstitial Ti atoms between TiS$_2$ layers. A subsequent study showed that the power-law of $\varrho(T)$ strongly varies with $x$ - hence with carrier concentration - and that the dominant scattering mechanism is phononic \cite{kli81}.     

Assuming a FL scenario, we quantify the strength of the electron-electron scattering by analyzing the curve of Fig. 1 using the FL expression $\varrho(T)=\varrho_0 + A T^2$. A linear fit of the curve in the 30-70 K region above the upturn yields $A=8.4 \times 10^{-3} \mu\Omega$cm K$^{-2}$, two orders of magnitude larger than typical values for transition metals \cite{ric68} and comparable to the values reported on the aforementioned TM oxides V$_2$O$_3$ \cite{mcw69}, Sr$_2$RuO$_4$ \cite{mae97} or LiV$_2$O$_4$ \cite{kon97}, which are established Fermi liquids. We conclude that CVS exhibits intermediate heavy-fermion properties. This conclusion is confirmed by the large electronic contribution to the specific heat, $\gamma$, obtained from a $C/T$ vs. $T^2$ plot (see Fig. 2). A linear fit yields $\gamma = 60 \pm 4$ mJ mol$^{-1}$ K$^{-2}$ and a Debye temperature $\Theta_D =294 \pm 5$ K \footnote{The statistical uncertainty in the $\gamma$ and $\beta$ values arises from the choice of the region of the $C/T$ vs. $T^2$ plot for the linear fit.}. Remarkably, Fig. 3 shows that the experimental $A$ and $\gamma$ values nicely fall onto the universal curve $A/\gamma^2 = 1.0 \times 10^{-5} \mu \Omega$ cm (mol K mJ$^{-1})^2$ found by Kadowaki and Woods for heavy fermions \cite{kad86}\footnote{We consider the $\gamma$ value per mole of planar V ions because each formula unit contains two V ions in the conducting VS$_2$ planes.} in the regime of intermediate mass renormalization, $m^{\ast}/m = 30$.    

\begin{figure}[ht]
\includegraphics[width=\columnwidth]{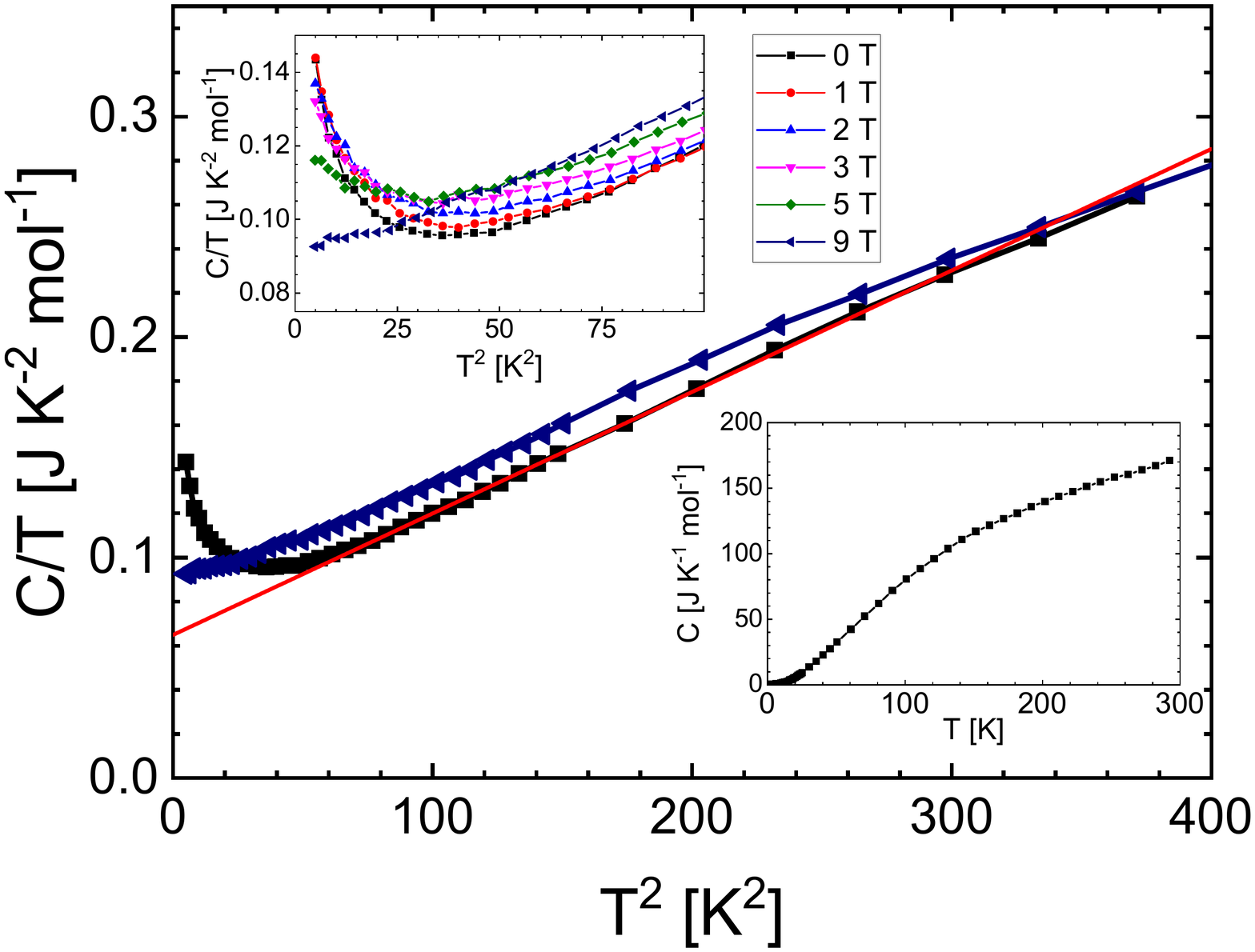}
\caption{(Color online) Temperature and magnetic field dependence of the constant-pressure specific heat, $C$, measured on a bunch of \CuVS~ single-crystals. The main panel shows the zero-field and 9 T curves and a fit of the former curve (red solid line) in the 10-20 K region using the linear dependence, $C/T = \gamma + \beta T^2$, expected from band theory at low temperatures. Note a marked upturn of the zero-field curve suppressed by the 9 T field. Top inset: enlarged view of the low-temperature behavior of the $C/T$ vs. $T^2$ curves at various fields in the 0-9 T range. Bottom inset: zero-field $C$ vs. $T$ curve in the whole temperature range measured.}
\label{fig2}
\end{figure}

\begin{figure}[ht]
\includegraphics[width=\columnwidth]{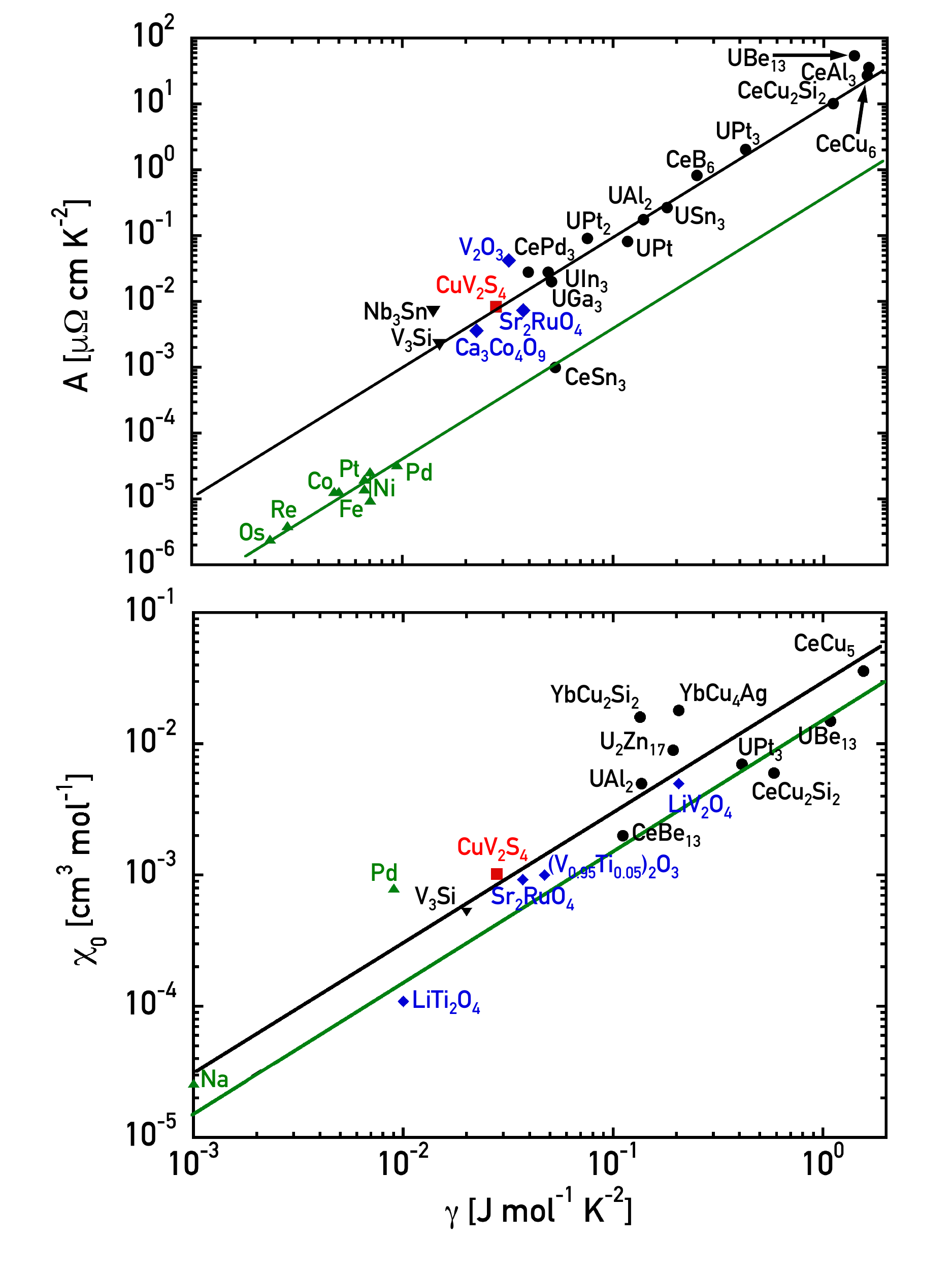}
\caption{Upper panel: Kadowaki-Woods plot of the $T^2$ resistivity coefficient, $A$, vs. Sommerfeld constant, $\gamma$, where the present result for \CuVS~ (red square) is compared with previous data for heavy-fermions (black circles) \cite{kad86}, $3d$ TM oxides (blue diamonds) \cite{mcw69,mae97}, A15 superconductors (black triangles) \cite{miy69} and simple metals (green triangles) \cite{ric68}. Black and green solid lines indicate the universal predictions including \cite{miy69} or neglecting \cite{ric68}) many-body effects, respectively. Lower panel: the same as above for the Sommerfeld-Wilson plot of the Pauli susceptibility, $\chi_0$ vs. $\gamma$. Black and green solid lines indicate the universal predictions for free-electron systems and heavy fermions \cite{lee86}, respectively. The $\chi_0$ value for \CuVS~ (red square) is taken from \cite{kle11}. Data include heavy-fermions (black circles) \cite{lee86}, $3d$-metal oxides (blue diamonds) \cite{men70,mae97,joh99}, A15 superconductors (black triangles) \cite{lee86} and simple metals (green triangles) \cite{lee86}.}
\label{fig3}
\end{figure}

The scenario of correlated metal is confirmed by the universal prediction for the Sommerfeld-Wilson ratio, $R_W=\frac{4}{3}(\frac{\pi k_B}{g \mu_B})^2\frac{\chi_0}{\gamma}$, where $g$ is the gyromagnetic ratio of the electron, $\mu_B$ is the Bohr magneton and $\chi_0$ is the Pauli susceptibility. Within a FL picture, both $\chi_0$ and $\gamma$ are proportional to the density of states at the Fermi level, hence $R_W = 1$. On the other hand, in the Kondo impurity problem, $R_W$ increases up to 2 according to a robust result of renormalization group theory \cite{wil75b,noz80}. In our case, by taking the experimental value $\chi_0=1.02 \times 10^{-3}$ cm$^3$ mol$^{-1}$ from \cite{kle11}, we obtain $R_W=2$ within the experimental error (see Fig. 3), which surprisingly suggests that CVS is a correlated Kondo metal. Further Kondo signatures are the resistivity upturn at low temperature and a similar upturn of the $C/T$ curve at $\sim$6 K, both suppressed by a 9 T field (see Figs. 1-2). The $\sim 2\%$ increase of the resistivity is comparable with that of AuFe \cite{for70} and about ten times smaller than that of CuMn \cite{hae76}. 

\begin{figure}[ht]
\includegraphics[width=\columnwidth]{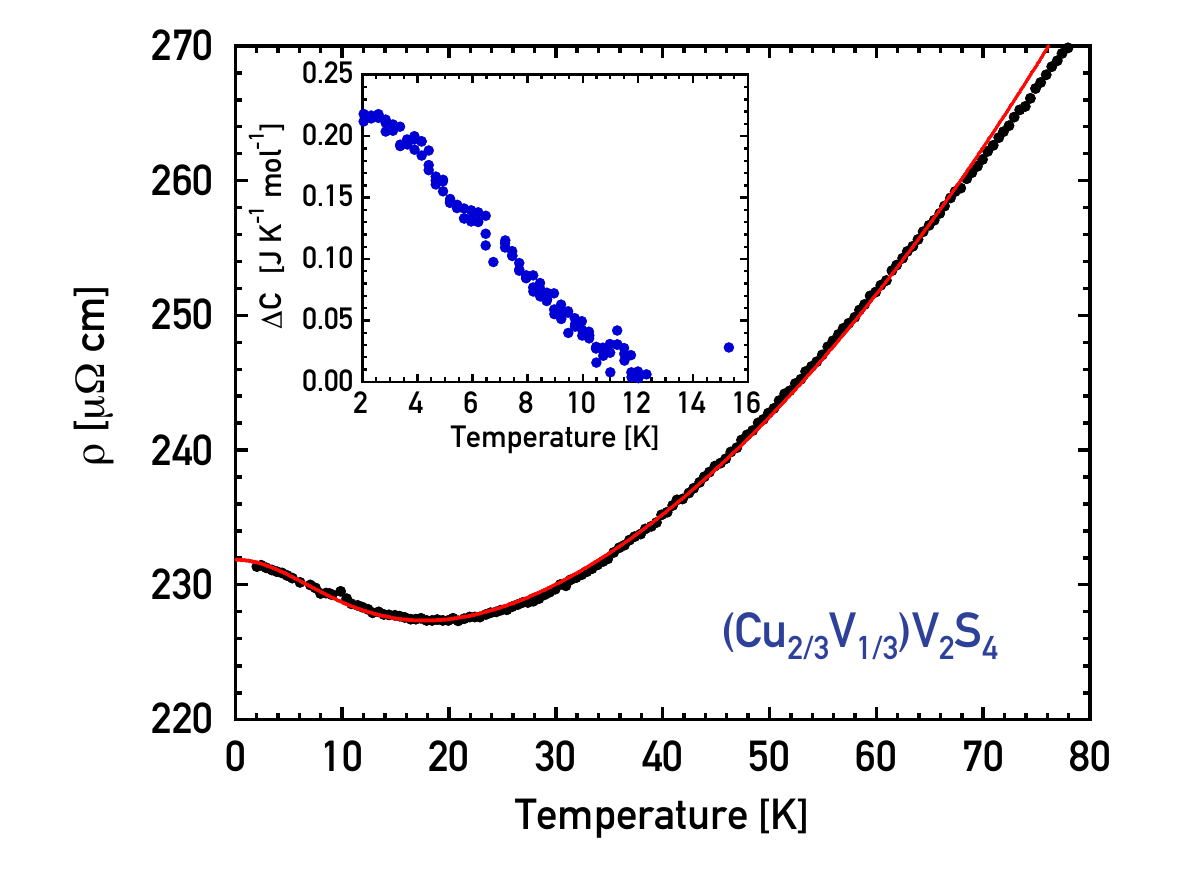}
\caption{(Color online) Comparison between the experimental resistivity at low temperatures (full black circles) and Nagaoka's prediction \cite{nag65} for a dilute Kondo model (red solid line), described by eq. (2). Top inset: temperature dependence of the excess specific heat, $\Delta C$, extracted from the linear fit of the $C/T$ vs $T^2$ plot in Fig. 2, as explained in the text. The linear dependence of $\Delta C$ is again consistent with Nagaosa's theory in the low-temperature limit (see text).}
\label{fig4}
\end{figure}

In order to check the validity of the Kondo picture, we analyze the temperature dependence of the upturn of the resistivity in Fig. 1. In the limit of diluted impurities, following the well-established result of Kondo's perturbative theory \cite{kon64} confirmed by subsequent studies that include a self-consistent treatment by Nagaoka \cite{nag65}, one finds that the excess resistivity $\Delta \varrho(T)$ describing the upturn diverges logarithmically as:

\begin{equation}
\Delta \varrho(T) = - c \frac{3 \pi}{16} \frac{m\mbox{*}}{ne^2} \frac{J}{\hbar} {\rm ln}\frac{T}{0.68 T_K}
\end{equation}

where $c$ is the impurity concentration, $m\mbox{*}$ and $n$ are the effective mass and density of conduction electrons, respectively and $J$ is the exchange energy between conduction electron and magnetic impurity. $J$ is related to the characteristic Kondo energy, $T_K$, via the BCS-like expressions $\Delta_0=1.14 k_B T_K$ and $\Delta_0=D{\rm exp}[-1/Jg(\epsilon_F)]$, where $\Delta_0$ is the zero-temperature Kondo energy, $D$ is the bandwidth and $g(\epsilon_F)$ is the density of states at the Fermi level, $\epsilon_F$. This perturbative result is valid only in the high temperature limit, $T \gg T_K$. In the opposite limit, various authors found that the logarithmic divergence is removed as electron-electron interactions become unimportant \cite{nag65,and70,noz74,wil75b}. The system then recovers a FL regime that manifests itself as a saturation of the resistivity described by a Lorentzian dependence \cite{nag65}:

\begin{equation}
\Delta \varrho(T)=\Delta \varrho(0)\left[1 + \frac{\pi^2}{3} \left(\frac{k_B T}{\Delta} \right)^2\right]^{-1}
\end{equation}

A straightforward data fit shows that the $\varrho(T)$ curve of Fig. 1 is very well explained in a wide 2-70 K range by adding the $\Delta \varrho(T)$ term of eq. (2) to the Fermi liquid contribution $\varrho_0 + A T^2$. In Fig. 4, we plot the $\varrho(T)$ curve together with the best fit that yields $T_K = 27.6 \pm 0.2$ K, consistent with the rule of thumb that $T_K$ corresponds to the resistivity minimum. The high quality of the agreement does not even require using eq. (1), instead of eq. (2), in the high temperature region $T \gg T_K$ where, in principle, eq. (1) would be more suitable. Note that the saturation of the resistivity upturn at low temperature, visible in Fig. 4 and explained quantitatively by eq. (2), rules out a weak-localization scenario, according to which the resistivity should diverge in the $T \rightarrow 0$ K limit. Other Kondo systems like CuFe and CuMn display a similar saturation of the resistivity upturn \cite{riz74}. 

The analysis of the specific heat data of Fig. 2 further supports a picture of Kondo system. We determine the Kondo contribution, $\Delta C(T)$, associated with the upturn of the experimental $C/T$ vs $T^2$ curve, by substracting from this curve the conventional dependence $C_0/T = \gamma + \beta T^2$ that describes well the data above the upturn (see linear fit in Fig. 2). Fig. 4 shows that $\Delta C(T)$ displays a linear increase with decreasing temperature which levels off at $\sim$2 K. This behavior is again consistent with Nagaoka's prediction \cite{nag65} of domelike dependence of $\Delta C(T)$ for a dilute Kondo system. This dependence, which has no analytic form, is approximated by a linear dependence $\Delta C(T) = - K (T - T_K )$ for $T \sim T_K$, where $K$ is a positive constant, and by $C(T) \sim T$ in the zero-temperature limit. In the present case, according to eq. (3), the linear behavior of Fig. 4 yields an estimate of $T_K \approx 12$ K, somehow lower than the $T_K$ estimated from the resistivity. The discrepancy is ascribed to an inaccurate determination of the electronic specific heat from the dominant lattice contribution or to the limitations of current nonperturbative theories of the Kondo effect \cite{and83,tsv83}. A further analysis of this point goes beyond the scope of the present paper.

We should compare the present upturns of the resistivity and of the specific heat at low temperature with similar features previously reported in $d$- \cite{nis97,lue99} and $f$-compounds \cite{ste84}. These upturns have been the object of controversial interpretations and an alternative scenario of Schottky anomaly has been put forward for compounds like the Heusler alloy Fe$_2$VAl \cite{lue99}. We verified this alternative scenario by analyzing the field-dependence of the excess specific heat, $\Delta C$, determined as above. According to a Schottky multi-level model \cite{fal71}, the low-temperature dependence of $\Delta C$ is expected to exhibit a hump which shifts toward high temperatures with increasing field. This prediction differs markedly from the behavior of Fig. 2. The model further predicts a hump for the field dependence of $\Delta C$ at constant temperature as well, for the magnetic contribution $\Delta C$ depends on the adimensional variable $x = \frac{g \mu_B H}{k_B T}$. Again, the behavior of Fig. 5 differs qualitatively from this prediction: at 2 K, the $\Delta C$ curve exhibits a linear decrease with field while, at this temperature and for a $S = 1/2$ ($S = 1$) impurity, a hump is predicted at $H = 3.5$ T ($2.5$ T). Thus, we can rule out the Schottky scenario in our case.             

We finally propose an intuitive explanation of the Kondo scenario suggested here. This scenario would be unique for a $d$-system, as the Kondo effect requires either localized magnetic impurities or narrow-bands characteristic of $f$- rather than $d$-systems \cite{lue99}. The peculiarity of the present $d$-system is the stacking of metallic VS$_2$ layers with (Cu,V)S chains containing one magnetic V$^{3+}$ ion every two nonmagnetic Cu$^{1+}$ ions (see Fig. 5). We argue that the $3d$ electrons of these ions are localized due to the reduced dimension of the chain, so the V$^{3+}$ ions behave as diluted $S = 1$ magnetic impurities interacting with the metallic layers. This explanation is supported by a previous analysis of the magnetic susceptibility \cite{kle11}, which indicates that only the V$^{3+}$ ions in the chain contribute to the Curie behavior. Suitable probes, such as angular resolved photoemission spectroscopy, may confirm experimentally the above scenario of onedimensional band.

\begin{figure}[ht]
\includegraphics[width=\columnwidth]{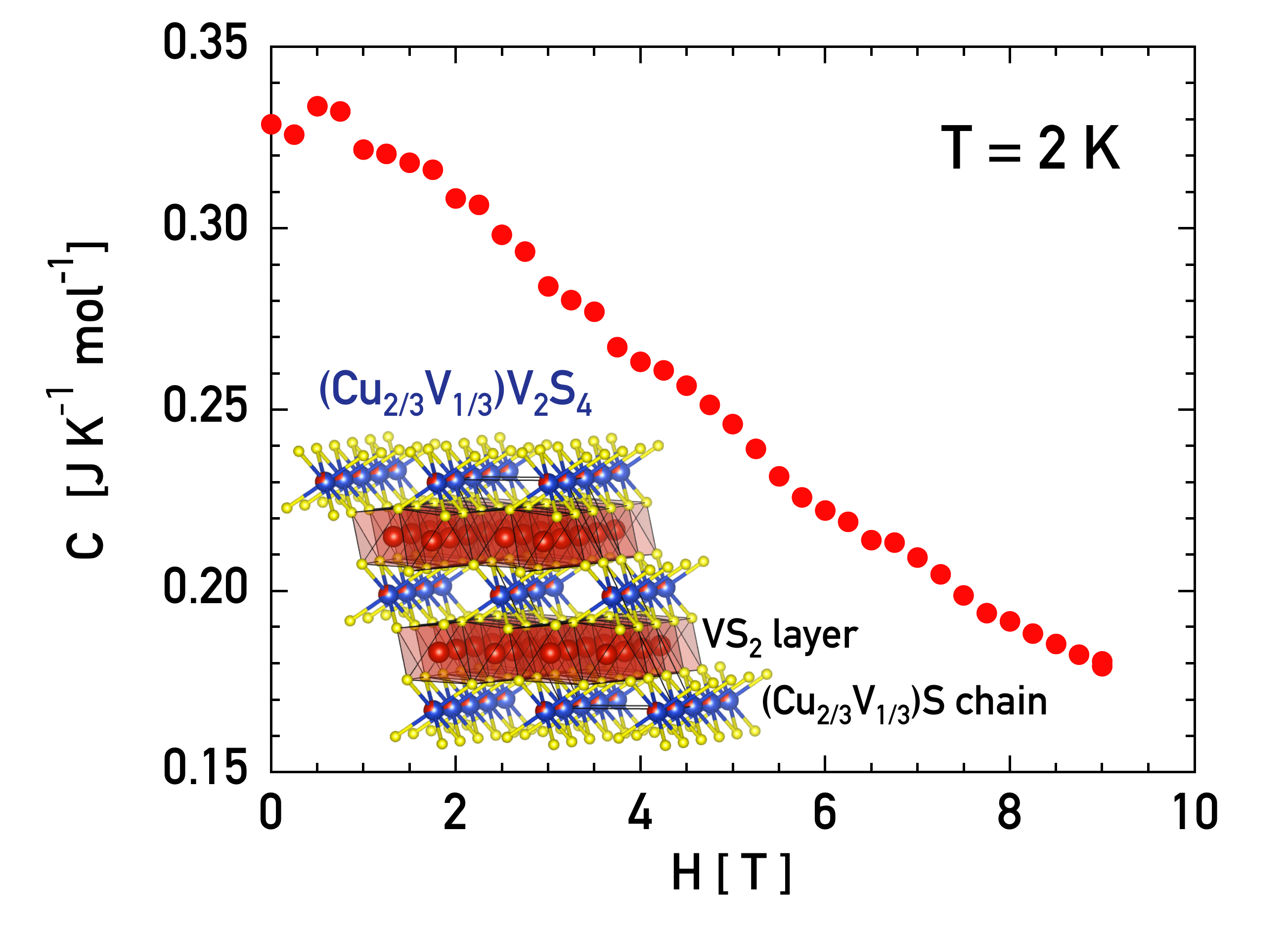}
\caption{(Color online) Magnetic field dependence of the specific heat of \CuVS~ at 2 K. Inset: schematic crystal structure showing the stacking of 1T-type VS$_2$ layers (red octahedra) and of (Cu$_{2/3}$V$_{1/3}$)S chains. Yellow and red/blue spheres indicate the S and Cu/V atoms, respectively.}
\label{fig5}
\end{figure}

In conclusion, we have given experimental evidence of a FL to NFL crossover induced by a Kondo transition in the twodimensional system \CuVS~(CVS). To the best of our knowledge, this is a unique example of correlated electron system exhibiting the signatures of both heavy-fermions in a broad 30-70 K temperature range and of Kondo transition. We therefore think that CVS constitutes a playground for studying the above crossover in a regime of intermediate electronic correlations accessible to first-principles calculations. For instance, we envisage that CVS and related compounds may constitute a simple realization of the two-channel Kondo state \cite{noz80,zaw80,pot07}. Indeed, if the coupling between layers was sufficiently weak, each metallic layer would screen independently the impurity spin in the chain site. The verification of this scenario awaits further studies at very low temperatures.

\begin{acknowledgments}
The authors acknowledge stimulating discussions with M. Calandra, M. Casula, P. Coleman and M. Hellgren.
\end{acknowledgments} 

\bibliographystyle{apsrev}
\bibliography{kondo}

\end{document}